\RequirePackage{lineno}   
\documentclass[aps,prl,twocolumn,superscriptaddress,showpacs,preprintnumbers,amsmath,amssymb,nofootinbib]{revtex4}
\usepackage[caption=false]{subfig}
\usepackage{graphicx} 
\usepackage{dcolumn}  
\usepackage{color}
\usepackage[titletoc]{appendix}
\graphicspath{{ps}}

\begin{document}



\title{Observation of $\tau^- \rightarrow \pi^- \nu_{\tau} e^+ e^-$ and search for $\tau^- \rightarrow \pi^- \nu_{\tau} \mu^+ \mu^-$}

\noaffiliation
\affiliation{University of the Basque Country UPV/EHU, 48080 Bilbao}
\affiliation{Beihang University, Beijing 100191}
\affiliation{Brookhaven National Laboratory, Upton, New York 11973}
\affiliation{Budker Institute of Nuclear Physics SB RAS, Novosibirsk 630090}
\affiliation{Faculty of Mathematics and Physics, Charles University, 121 16 Prague}
\affiliation{Chonnam National University, Kwangju 660-701}
\affiliation{University of Cincinnati, Cincinnati, Ohio 45221}
\affiliation{Deutsches Elektronen--Synchrotron, 22607 Hamburg}
\affiliation{Duke University, Durham, North Carolina 27708}
\affiliation{Key Laboratory of Nuclear Physics and Ion-beam Application (MOE) and Institute of Modern Physics, Fudan University, Shanghai 200443}
\affiliation{Justus-Liebig-Universit\"at Gie\ss{}en, 35392 Gie\ss{}en}
\affiliation{SOKENDAI (The Graduate University for Advanced Studies), Hayama 240-0193}
\affiliation{Hanyang University, Seoul 133-791}
\affiliation{University of Hawaii, Honolulu, Hawaii 96822}
\affiliation{High Energy Accelerator Research Organization (KEK), Tsukuba 305-0801}
\affiliation{J-PARC Branch, KEK Theory Center, High Energy Accelerator Research Organization (KEK), Tsukuba 305-0801}
\affiliation{Forschungszentrum J\"{u}lich, 52425 J\"{u}lich}
\affiliation{IKERBASQUE, Basque Foundation for Science, 48013 Bilbao}
\affiliation{Indian Institute of Science Education and Research Mohali, SAS Nagar, 140306}
\affiliation{Indian Institute of Technology Bhubaneswar, Satya Nagar 751007}
\affiliation{Indian Institute of Technology Guwahati, Assam 781039}
\affiliation{Indian Institute of Technology Hyderabad, Telangana 502285}
\affiliation{Indian Institute of Technology Madras, Chennai 600036}
\affiliation{Indiana University, Bloomington, Indiana 47408}
\affiliation{Institute of High Energy Physics, Chinese Academy of Sciences, Beijing 100049}
\affiliation{Institute of High Energy Physics, Vienna 1050}
\affiliation{Institute for High Energy Physics, Protvino 142281}
\affiliation{INFN - Sezione di Napoli, 80126 Napoli}
\affiliation{INFN - Sezione di Torino, 10125 Torino}
\affiliation{Advanced Science Research Center, Japan Atomic Energy Agency, Naka 319-1195}
\affiliation{J. Stefan Institute, 1000 Ljubljana}
\affiliation{Institut f\"ur Experimentelle Teilchenphysik, Karlsruher Institut f\"ur Technologie, 76131 Karlsruhe}
\affiliation{Kennesaw State University, Kennesaw, Georgia 30144}
\affiliation{King Abdulaziz City for Science and Technology, Riyadh 11442}
\affiliation{Department of Physics, Faculty of Science, King Abdulaziz University, Jeddah 21589}
\affiliation{Kitasato University, Sagamihara 252-0373}
\affiliation{Korea Institute of Science and Technology Information, Daejeon 305-806}
\affiliation{Korea University, Seoul 136-713}
\affiliation{Kyoto University, Kyoto 606-8502}
\affiliation{Kyungpook National University, Daegu 702-701}
\affiliation{LAL, Univ. Paris-Sud, CNRS/IN2P3, Universit\'{e} Paris-Saclay, Orsay}
\affiliation{\'Ecole Polytechnique F\'ed\'erale de Lausanne (EPFL), Lausanne 1015}
\affiliation{P.N. Lebedev Physical Institute of the Russian Academy of Sciences, Moscow 119991}
\affiliation{Liaoning Normal University, Dalian 116029}
\affiliation{Faculty of Mathematics and Physics, University of Ljubljana, 1000 Ljubljana}
\affiliation{Ludwig Maximilians University, 80539 Munich}
\affiliation{Luther College, Decorah, Iowa 52101}
\affiliation{Malaviya National Institute of Technology Jaipur, Jaipur 302017}
\affiliation{University of Malaya, 50603 Kuala Lumpur}
\affiliation{University of Maribor, 2000 Maribor}
\affiliation{Max-Planck-Institut f\"ur Physik, 80805 M\"unchen}
\affiliation{School of Physics, University of Melbourne, Victoria 3010}
\affiliation{University of Mississippi, University, Mississippi 38677}
\affiliation{University of Miyazaki, Miyazaki 889-2192}
\affiliation{Moscow Physical Engineering Institute, Moscow 115409}
\affiliation{Moscow Institute of Physics and Technology, Moscow Region 141700}
\affiliation{Graduate School of Science, Nagoya University, Nagoya 464-8602}
\affiliation{Kobayashi-Maskawa Institute, Nagoya University, Nagoya 464-8602}
\affiliation{Universit\`{a} di Napoli Federico II, 80055 Napoli}
\affiliation{Nara Women's University, Nara 630-8506}
\affiliation{National Central University, Chung-li 32054}
\affiliation{National United University, Miao Li 36003}
\affiliation{Department of Physics, National Taiwan University, Taipei 10617}
\affiliation{H. Niewodniczanski Institute of Nuclear Physics, Krakow 31-342}
\affiliation{Nippon Dental University, Niigata 951-8580}
\affiliation{Niigata University, Niigata 950-2181}
\affiliation{University of Nova Gorica, 5000 Nova Gorica}
\affiliation{Novosibirsk State University, Novosibirsk 630090}
\affiliation{Osaka City University, Osaka 558-8585}
\affiliation{Pacific Northwest National Laboratory, Richland, Washington 99352}
\affiliation{Panjab University, Chandigarh 160014}
\affiliation{Peking University, Beijing 100871}
\affiliation{University of Pittsburgh, Pittsburgh, Pennsylvania 15260}
\affiliation{Punjab Agricultural University, Ludhiana 141004}
\affiliation{Theoretical Research Division, Nishina Center, RIKEN, Saitama 351-0198}
\affiliation{University of Science and Technology of China, Hefei 230026}
\affiliation{Seoul National University, Seoul 151-742}
\affiliation{Showa Pharmaceutical University, Tokyo 194-8543}
\affiliation{Soongsil University, Seoul 156-743}
\affiliation{Stefan Meyer Institute for Subatomic Physics, Vienna 1090}
\affiliation{Sungkyunkwan University, Suwon 440-746}
\affiliation{Department of Physics, Faculty of Science, University of Tabuk, Tabuk 71451}
\affiliation{Tata Institute of Fundamental Research, Mumbai 400005}
\affiliation{Department of Physics, Technische Universit\"at M\"unchen, 85748 Garching}
\affiliation{Toho University, Funabashi 274-8510}
\affiliation{Department of Physics, Tohoku University, Sendai 980-8578}
\affiliation{Earthquake Research Institute, University of Tokyo, Tokyo 113-0032}
\affiliation{Department of Physics, University of Tokyo, Tokyo 113-0033}
\affiliation{Tokyo Institute of Technology, Tokyo 152-8550}
\affiliation{Tokyo Metropolitan University, Tokyo 192-0397}
\affiliation{Virginia Polytechnic Institute and State University, Blacksburg, Virginia 24061}
\affiliation{Wayne State University, Detroit, Michigan 48202}
\affiliation{Yamagata University, Yamagata 990-8560}
\affiliation{Yonsei University, Seoul 120-749}
  \author{Y.~Jin}\affiliation{Department of Physics, University of Tokyo, Tokyo 113-0033} 
  \author{H.~Aihara}\affiliation{Department of Physics, University of Tokyo, Tokyo 113-0033} 
  \author{D.~Epifanov}\affiliation{Budker Institute of Nuclear Physics SB RAS, Novosibirsk 630090}\affiliation{Novosibirsk State University, Novosibirsk 630090} 
  \author{I.~Adachi}\affiliation{High Energy Accelerator Research Organization (KEK), Tsukuba 305-0801}\affiliation{SOKENDAI (The Graduate University for Advanced Studies), Hayama 240-0193} 
  \author{S.~Al~Said}\affiliation{Department of Physics, Faculty of Science, University of Tabuk, Tabuk 71451}\affiliation{Department of Physics, Faculty of Science, King Abdulaziz University, Jeddah 21589} 
  \author{D.~M.~Asner}\affiliation{Brookhaven National Laboratory, Upton, New York 11973} 
  \author{V.~Aulchenko}\affiliation{Budker Institute of Nuclear Physics SB RAS, Novosibirsk 630090}\affiliation{Novosibirsk State University, Novosibirsk 630090} 
  \author{T.~Aushev}\affiliation{Moscow Institute of Physics and Technology, Moscow Region 141700} 
  \author{R.~Ayad}\affiliation{Department of Physics, Faculty of Science, University of Tabuk, Tabuk 71451} 
  \author{V.~Babu}\affiliation{Deutsches Elektronen--Synchrotron, 22607 Hamburg} 
% ...
\collaboration{The Belle Collaboration}


\begin{abstract}
We present the first measurements of branching fractions of rare tau-lepton decays, 
$\tau^- \rightarrow \pi^- \nu_{\tau} \ell^+ \ell^-$ ($\ell = e$ or $\mu$), using a data sample corresponding to 562 fb$^{-1}$ collected at a center-of-mass energy of 10.58 GeV with the Belle detector at the KEKB asymmetric-energy 
$e^+ e^-$ collider. The $\tau^- \rightarrow \pi^- \nu_\tau e^+ e^-$ decay is observed for the first time 
with 7.0$\sigma$ significance. 
The partial branching fraction determined by the structure-dependent mechanisms mediated 
by either a vector or an axial-vector current for the mass region $M_{\pi e e}>1.05$~GeV/$c^2$ is measured 
to be $\mathcal{B}(\tau^-\rightarrow \pi^- \nu_\tau e^+ e^-)[M_{\pi^- e^+ e^-}>1.05~{\rm GeV}/c^2] = (5.90 \pm 0.53 \pm 0.85 \pm 0.11) \times 10^{-6}$, 
where the first uncertainty is statistical, the second is systematic, and the third is due 
to the model dependence. In the full phase space, due to the different detection efficiencies for 
the structure-dependent mechanisms mediated by axial-vector and vector 
currents, the branching fraction varies from 
$\mathcal{B}_{A}(\tau^-\rightarrow \pi^- \nu_\tau e^+ e^-) = (1.46 \pm 0.13 \pm 0.21) \times 10^{-5}$ to 
$\mathcal{B}_{V}(\tau^-\rightarrow \pi^- \nu_\tau e^+ e^-) = (3.01 \pm 0.27 \pm 0.43) \times 10^{-5}$, respectively.
An upper limit is set on the branching fraction of the $\tau^- \rightarrow \pi^- \nu_\tau \mu^+ \mu^-$ decay, 
$\mathcal{B}(\tau^-\rightarrow \pi^- \nu_\tau \mu^+ \mu^-) < 1.14 \times 10^{-5}$, at the 90\% confidence level. 
\end{abstract} 


\maketitle

\tighten

{\renewcommand{\thefootnote}{\fnsymbol{footnote}}}
\setcounter{footnote}{0}

\section{I. Introduction}
The hadronic final states of the tau-lepton decays provide a clean laboratory to study the dynamics of strong interactions in the energy region below the tau mass. 
The world's largest sample of tau leptons, collected with the Belle detector, allows measurement of rare tau decay branching fractions to a new level of sensitivity. 
We present the first measurements of branching fractions of rare tau-lepton decays $\tau^- \rightarrow \pi^- \nu_{\tau} \ell^+ \ell^-$ ($\ell = e$ or $\mu$)~\cite{CC}, 
whose theoretical predictions are calculated as $\mathcal{B}(\tau^- \rightarrow \pi^- \nu_{\tau} e^+ e^-) \in [1.4, 2.8]\times 10^{-5}$ and 
$\mathcal{B}(\tau^- \rightarrow \pi^- \nu_{\tau} \mu^+ \mu^-) \in [0.03, 1.0] \times 10^{-5}$~\cite{Roig}. 

\begin{figure}[!htbp]
	\includegraphics[width=0.5\textwidth]{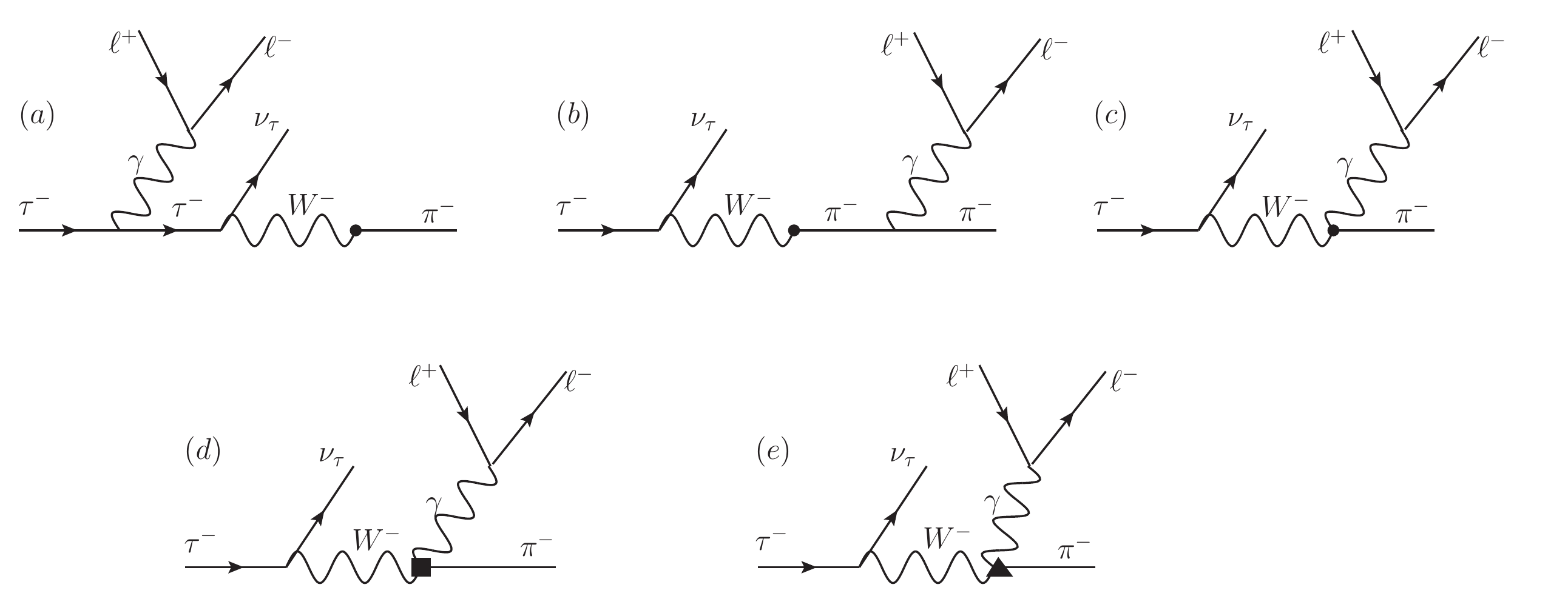}
	\caption{Feynman diagrams for the dominant amplitudes for $\tau^- \to \pi^- \nu_{\tau} \ell^+ \ell^-$~\cite{Roig}. The dot indicates known (fully data-driven) hadronization effects. The solid square (triangle) represents the structure-dependent contribution mediated by the vector (axial-vector) current.} \label{xm} 
\end{figure}

These decays are interesting because they involve the $\gamma^* W^* \pi$ vertex with two gauge bosons off their mass shells. The relevant amplitudes 
for the decays $\tau^- \rightarrow \pi^- \nu_\tau \ell^+ \ell^-$ can be written as a sum of five contributions, as shown in Fig.~\ref{xm}~\cite{explanation}. 
The diagrams (a), (b), and (c) are structure-independent while the diagram (d) and (e) are structure-dependent. 
In all amplitudes, the lepton pair is produced by a virtual photon.
In the case of a real photon \cite{realpho}, the $\gamma W \pi$ vertex plays an important role in determining the radiative corrections to the $\tau^- \rightarrow \pi^- \nu_\tau$ 
decay \cite{RC} and in the evaluation of the contributions of the hadronic light-by-light scattering to the muon anomalous magnetic moment \cite{g2}. The measured branching 
fractions of the decay modes of interest can be used to validate the Resonance Chiral Theory~\cite{RCT}, a semi-phenomenological approach to describe strong interactions at energies 
below the tau mass. In addition, rare decays of $\tau$ often serve as a probe of physics beyond the Standard Model~\cite{sterile}. 
Another important impact of a precise measurement of $\mathcal{B}(\tau^- \rightarrow \pi^- \nu_{\tau} \ell^+ \ell^-)$ is useful input for a reliable background estimation 
in searches for various lepton flavor and lepton number violating decays.

\section{II. Data Set and simulations}

In this paper, we report branching fraction measurements for the rare decays $ \tau^- \rightarrow \pi^- \nu_{\tau} e^+ e^-$ and $\tau^- \rightarrow \pi^- \nu_{\tau} \mu^+ \mu^-$ 
with data recorded at a center-of-mass (CM) energy of $\sqrt{s}=10.58$ GeV by the Belle experiment at the KEKB asymmetric-energy $e^+ e^-$ collider~\cite{KEKB}. The data used in 
this analysis correspond to an integrated luminosity of $562~{\rm fb}^{-1}$. The Belle detector is a general purpose large-solid-angle spectrometer consisting of a silicon vertex 
detector, a central drift chamber (CDC), an array of aerogel threshold Cherenkov counters (ACC), a barrel-like arrangement of time-of-flight scintillation counters (TOF), and an 
electromagnetic calorimeter (ECL) located inside a superconducting solenoid coil that provides a 1.5 T magnetic field. Outside the coil, an iron flux-return yoke is instrumented 
to detect $K^{0}_{L}$ mesons and to identify muons (KLM). A detailed description of the Belle detector can be found elsewhere~\cite{thebelledetector}. In this analysis, we use the 
data set collected with an inner configuration that comprises a 1.5 cm radius beampipe, a 4-layer silicon detector and a small-cell inner drift chamber. 

To optimize the event selection criteria and to determine the detection efficiency for the signal events, Monte Carlo (MC) samples are employed. The MC samples are generated 
using EVTGEN~\cite{EVTGEN} for hadronic processes, KKMC~\cite{KK} for fermion pairs, and AAFH~\cite{AAFH} for two-photon production of fermion pairs. The $\tau$ lepton decays 
are carried out by TAUOLA~\cite{TAUOLA}. The final-state radiation of charged particles is simulated by PHOTOS~\cite{PHOTOS}. For the signal modes 
$ \tau^- \rightarrow \pi^- \nu_{\tau} \ell^+ \ell^- $, we employ formulas given in Ref.~\cite{Roig} and implement them into the TAUOLA generator. 
For the $\tau^- \to \pi^- \pi^0 \nu_{\tau}$ mode, the largest background for the $\tau^- \to \pi^- \nu_{\tau} e^+ e^-$ signal, we use the pion form factor measured by the Belle 
collaboration~\cite{Hayashii}. For the $\pi^0$ decay, in addition to $\pi^0 \to \gamma \gamma$, the $\pi^0$ Dalitz decay, $\pi^0 \to \gamma e^+ e^-$ (1.17\%), is implemented via the pythia package~\cite{pythia}. The response of the detector is simulated by a GEANT3-based program~\cite{GEANT3}. The optimization of selection criteria 
using MC samples is implemented by maximizing $S/\sqrt{S+B}$, where $S$ ($B$) is the number of signal (background) events. 

\section{III. Event selection and reconstruction}

The selection proceeds in two stages, aimed at suppressing background processes while retaining a high efficiency for the decay mode of interest. At the first stage, we select 
$e^+ e^- \rightarrow \tau^+ \tau^-$ events and substantially reject background from the other processes that occur at $\Upsilon(4S)$. The second stage proceeds to select one 
$\tau$ decaying into the $\pi^- \nu_{\tau} \ell^+ \ell^-$ final state, where a vertex fit is performed to the  $\ell^+ \ell^-$ pair, and the other $\tau$ decaying into one-prong modes. 

\subsection{A. Selection of $\tau^{+}\tau^{-}$ events}

Events with four tracks and zero net charge are selected. The distance of closest approach of each charged particle to the interaction point (IP) must be less than 5 cm along the beam 
direction (the $z$ axis) and less than 1 cm in the transverse plane (the $x-y$ plane). Each charged particle must have transverse momentum larger than 0.1 GeV/$c$; and at least one particle 
must have transverse momentum larger than 0.5 GeV/$c$. The total energy deposited in the ECL must be less than 10 GeV. A vertex fit is performed using all charged particles to determine the 
position of the primary vertex of the event, which is required to be near the interaction point within 0.5 cm in the $x-y$ plane and 2.5 cm along the $z$ axis. The mean position of the interaction point 
itself is monitored using generic multi-track events in the same data taking period with an accuracy of 10 $\mu$m. Photons, reconstructed from the clusters in the ECL not associated with tracks, 
are selected with energy thresholds of 50 MeV in the barrel region ($32^\circ < \theta_{\gamma} <130^\circ$) and 100 MeV in the endcap region to eliminate beam-background photons. 

In order to reduce the remaining background from radiative Bhabha events, continuum $e^+ e^- \to q \bar{q}$ (where $q=u,d,s,c$) production, and two-photon processes, we require the following conditions. 
In the CM frame, the sum of the magnitudes of charged particle momenta must be less than 10 GeV/$c$, and its sum with photon energies must exceed 3 GeV. Opening angles for every pair of 
tracks must be less than $175^\circ$. A tau-pair event is accompanied by missing four-momentum due to neutrinos, defined as $p_{\rm miss}=p_{\rm init}-\sum_{i} p_{{\rm trk}, i} - \sum_{i} p_{\gamma, i}$, 
where $p_{\rm init}$ is the four-momentum of the colliding $e^+e^-$ beams, and $\sum_{i} p_{{\rm trk}, i}$ and $\sum_{i} p_{\gamma, i}$ are the sums of the four-momenta of charged particles (assumed to be pions) 
and photons, respectively. Only events satisfying $1~{\rm GeV}/c^2 < M_{\rm miss}<7~{\rm GeV}/c^2$ and $30^\circ < \theta_{\rm miss} <150^\circ$ criteria are selected, where $M_{\rm miss}$ and $\theta_{\rm miss}$ are 
the mass and polar angle of the missing four-momentum in the CM frame. 

We use the thrust variable to reduce $e^+ e^- \to q \bar{q}$ backgrounds and to provide two disjoint hemispheres for each event. The thrust $T$ is defined as the maximum value of 
${\sum_{i} |\vec{p}_{i} \cdot \vec{n}_{T}|} / {\sum_{i} |\vec{p}_{i}|}$ with respect to the thrust axis $\vec{n}_{T}$, 
where $\vec{p}_{i}$ is the CM momentum of the $i$th particle, either charged or neutral. We require the thrust $T$ to be in the range [0.85, 0.99]. 
Events are then divided into two hemispheres, in the CM frame, by the plane perpendicular to the thrust axis. We require three charged particles 
in the signal hemisphere and one charged particle in the other.

\subsection{B. Selection of $\tau^{-} \rightarrow \pi^{-} \nu_{\tau} e^+ e^-$ events}

To select $\tau^{-} \rightarrow \pi^{-} \nu_{\tau} e^+ e^-$ events, we require one charged pion, one electron, and one positron in the signal hemisphere. 
In order to identify a pion, pion (${\cal L}_\pi$) and kaon (${\cal L}_K$) likelihoods are constructed from the ACC response, the specific ionization ($dE/dx$) in the CDC, 
and the flight-time measurement in the TOF. A likelihood ratio ${\cal P}_{K/\pi}={\cal L}_K /({\cal L}_\pi +{\cal L}_K)$ is formed and we require that the pion track satisfies 
$\mathcal{P}_{K/\pi}<0.6$.  To select electrons, a likelihood ratio is required to satisfy ${\cal P}_{e}={\cal L}_{e}/({\cal L}_{e}+{\cal L}_{X})>0.5$, where the electron (${\cal L}_{e}$) 
and non-electron (${\cal L}_{X}$) likelihood functions include information on the $dE/dx$ in the CDC, the ratio of the energy of the cluster in the ECL 
to the momentum of the track, the transverse shape of the ECL shower, the matching of the track with the ECL cluster, and the signal amplitude in the ACC. A pion candidate is 
required to have ${\cal P}_{e}< 0.2$ and a momentum larger than 0.2 GeV/c in both the CM and lab frame. To recover from bremsstrahlung, four-momenta of photon candidates with 
a direction within 0.05 radians of the electron track and with an energy lower than that of the electron in the CM frame are added to the four-momentum of the electron track. 
In the signal hemisphere, after recovery, at most one photon is allowed and its energy must be below 300 MeV. 
To reduce $e^+ e^- \to q \bar{q}$ backgrounds, we use the variable
\begin{eqnarray}
x = \frac{2E_{\rm beam}E_{\pi e e}-m_{\tau}^2 \cdot c^4 -M_{\pi e e}^2 \cdot c^4}{2|p_{\tau}| \cdot |p_{\pi e e}| \cdot c^2}, 
\end{eqnarray}
where $E_{\rm beam}=\sqrt{s}/2$ is the beam energy, $E_{\pi e e}$, $p_{\pi e e}$, and $M_{\pi e e}$ are the energy, momentum, and invariant mass of the $\pi e e$ system in the CM 
frame. In the case of tau-lepton events, the variable $x$ is equal to $\cos\theta_{\tau-{\pi e e}}$, the cosine of the angle between the momentum of the $\tau$ and that of the 
$\pi e e$ system, assuming a massless tau neutrino. Therefore, the absolute value of $x$ is required not to exceed one. 

The main background to $\tau^{-} \rightarrow \pi^{-} \nu_{\tau} e^+ e^-$ is from $\tau^- \rightarrow \pi^- \pi^0 \nu_\tau$ decays, which have the largest branching fraction among 
all tau-lepton decay modes. The mode $\tau^- \rightarrow \pi^- \pi^0 \nu_\tau$ has a similar final state if one of the photons from $\pi^0 \to \gamma \gamma$ converts into 
$e^+ e^-$ pair in the detector or if the $\pi^0$ decays into its Dalitz mode ($\pi^0 \to \gamma e^+ e^-$). To suppress background due to external photon conversion, 
the transverse position of the $e^+ e^-$ vertex ($R_{xy}$) with respect to IP must be less than 1.2 cm, and the longitudinal position of the $e^+ e^-$ vertex must be in the 
range [$-1$ cm, 1.5 cm].  
In addition, we veto $\pi^0$ Dalitz decays using the photon that survives the aforementioned requirement and require the invariant mass of the $e^+ e^- \gamma$ system, 
$M_{e e \gamma}$, to fall outside of the range 110 MeV/$c^2$ $< M_{e e \gamma} <$165 MeV/$c^2$ [$\pm 3.8$ standard deviation ($\sigma$)]. 

Even after these requirements, the dominant background is still from $\tau^- \rightarrow \pi^- \pi^0 \nu_\tau$, and its reliable estimation is essential for 
the signal extraction. The $M_{\pi e e}$ range from 0 to 1 GeV/$c^2$ is chosen as a control region to validate the MC, where there is a large background from the
$\tau^- \rightarrow \rho^- (\to \pi^- \pi^0) \nu_\tau$ contribution, while we use the higher mass region, 1.05 GeV/$c^2$ $<M_{\pi e e}<$ 1.8 GeV/$c^2$, as the signal region. 
According to MC simulation the signal region we choose is efficient for the events with the structure-dependent mechanism (see Fig.~\ref{xm} (d),(e)), 
while the structure-independent signal events (see Fig.~\ref{xm} (a)-(c)) are located mostly in the region with $M_{\pi e e}<$ 1.05~GeV/$c^2$. As a result, this study is sensitive only to the structure-dependent contributions to the branching fraction of the $\tau^{-} \rightarrow \pi^{-} \nu_{\tau} e^+ e^-$ decay.
The signal region is blinded until we finalize the selection conditions and the corrections to the simulated samples. Several corrections, such as the efficiencies of the 
particle identification, tracking, and $\pi^0$ reconstruction are applied to the MC samples. 

\begin{figure}[!htbp]
	\includegraphics[width=5cm, height=4cm]{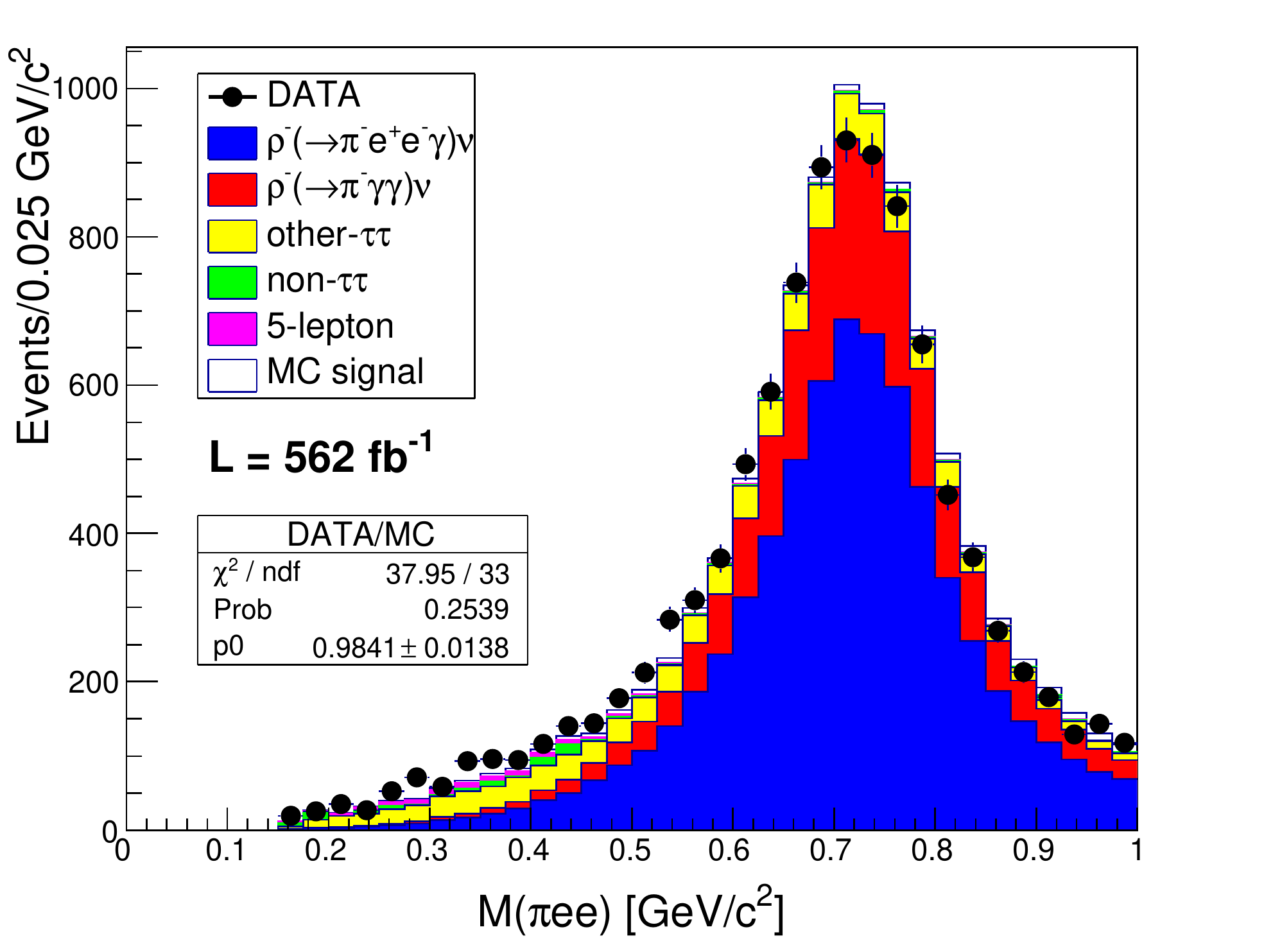}
	\caption{Distribution of $ \pi e^+ e^-$ invariant mass ($M_{\pi ee}$) in the control region for the $\tau^- \to \pi^- \nu_{\tau} e^+ e^-$ candidates. Circles with statistical 
                 error bars represent the experimental data. The histograms show the expectations from simulation for the $\tau^- \to \pi^- \nu_{\tau} e^+ e^-$ signal (white) 
                 and various background components (other colors, as explained in the legend). The MC samples are corrected for known efficiency differences between data and 
                 simulation, and are normalized to the luminosity of the data sample. 
	}
	\label{side1}
\end{figure}

Figure~\ref{side1} shows the $M_{\pi e e}$~distribution in the control region, where yields from data and the expected background estimated from MC simulation agree; 
10243 events are observed in the control region with a background expectation of 10083$\pm$504 events, where the uncertainty includes all systematic uncertainties discussed 
in section IV~A. The MC result shows that this control region is accurately described by only background events within 1$\sigma$, i.e., we don't observe a notable contribution from the structure-independent signal events in the control region.

\subsection{C. Selection of $\tau^{-} \rightarrow \pi^{-} \nu_{\tau} \mu^+ \mu^-$ events}

To select $\tau^{-} \rightarrow \pi^{-} \nu_{\tau} \mu^+ \mu^-$ events, we require one charged pion and two oppositely-charged muons in the signal hemisphere. 
The sum of energies of all photons in the signal hemisphere is required to be less than 300 MeV and less than six photons should be reconstructed in both hemispheres. 
This requirement is more stringent than in section III. B, since photon emission is negligible in this mode. The pion candidate is required to satisfy 
$\mathcal{P}_{K/\pi}<0.6$ (not kaon) and ${\cal P}_{e}<0.8$ (not electron). To select muon candidates, we employ a likelihood ratio 
${\cal P}_\mu={\cal L}_\mu /({\cal L}_\mu +{\cal L}_\pi +{\cal L}_K)$, where ${\cal L}_\mu$, ${\cal L}_\pi$, and ${\cal L}_K$ are likelihood functions. 
The likelihood for a muon, ${\cal L}_\mu$ is calculated from two variables: the difference between the penetration lengths determined from the momentum of the particle and measured 
by the KLM, and the $\chi^2$ of the KLM hits with respect to the track extrapolated from the CDC to the KLM. A stringent condition ${\cal P}_\mu>$ 0.97 is imposed upon muon candidates 
to suppress background events from $\tau^- \rightarrow \pi^- \pi^+ \pi^- \nu_\tau$, and the transverse momenta of the muons are required to exceed 720 MeV/$c$ to ensure that 
they reach the KLM detector. Furthermore, muon candidates are required to have ${\cal P}_{K/\pi}< 0.8$ to suppress the $\tau^- \rightarrow K^- \pi^+ \pi^- \nu_\tau$ mode. 
To suppress background from hadronic processes, the event is required to have a thrust value larger than 0.9. In addition, the pseudomass~\cite{pseudo} of the $\pi \mu \mu$ system, 
defined as 
\begin{eqnarray}
m^*=\biggl[ 2 \cdot E_{\pi \mu \mu}(E_{\rm beam} -E_{\pi \mu \mu}) / c^4+M^2_{\pi \mu \mu}    \nonumber  \\
-2 \cdot |p_{\pi \mu \mu}| \cdot (E_{\rm beam}-E_{\pi \mu \mu}) /c^3 \biggr]^\frac12,
\end{eqnarray}
is required to be less than 1.8 GeV/$c^2$, where $E_{\pi \mu \mu}$, $p_{\pi \mu \mu}$, and $M_{\pi \mu \mu}$ are the energy, momentum, and invariant mass of the $\pi \mu \mu$ system, 
respectively. A loose requirement on the invariant mass of the $\mu^+\mu^-$ system, $M(\mu^+ \mu^-)< 0.85$ GeV/$c^2$, is applied to further reduce this background. 

The remaining background to this decay mode is dominated by the decay $\tau^- \rightarrow \pi^- \pi^+ \pi^- (\pi^0) \nu_{\tau}$, where two charged pions are misidentified as muons. 
This misidentification happens mainly due to pion decays in flight. Therefore, many of these misidentified tracks do not originate from the IP. We define the signal region as 
$R_{xy}<$0.15 cm and the control region as $R_{xy}>$ 0.20 cm, where $R_{xy}$ is the radial position of the reconstructed $\mu^+\mu^-$ vertex with respect to the IP.

\begin{figure}[!htbp]
	\includegraphics[width=5cm, height=4cm]{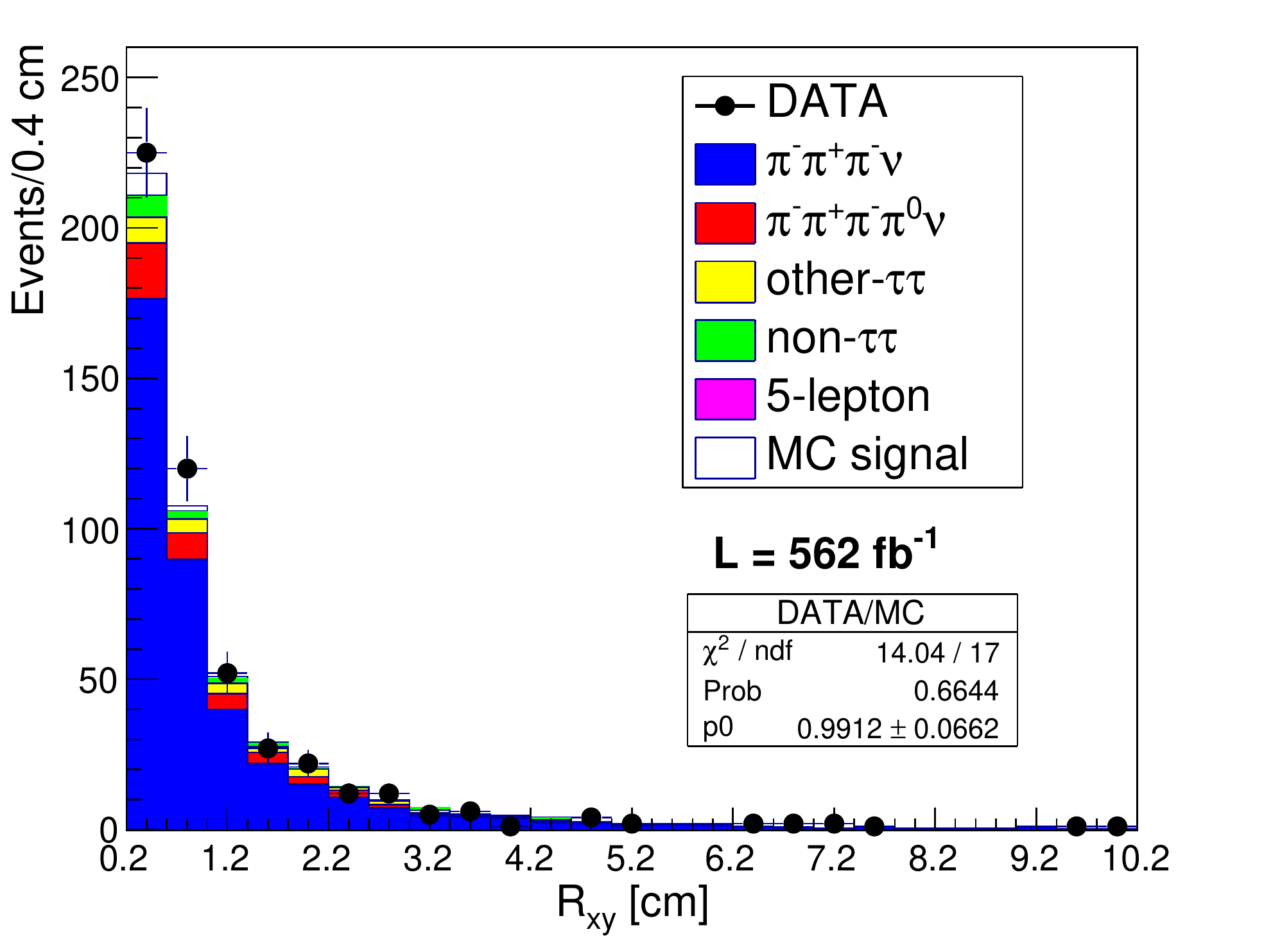}
	\caption{Distribution of transverse position of the $\mu^+ \mu^-$ vertex $(R_{xy})$ in the control region for the $\tau^- \rightarrow \pi^- \nu_{\tau} \mu^+ \mu^-$ 
                 candidates. Circles with statistical error bars represent the experimental data. The histograms show the expectations from MC simulation for 
                 the $\tau^- \rightarrow \pi^- \nu_{\tau} \mu^+ \mu^-$ signal (white) and various background components (other colors, as explained in the legend). 
                 See the text for more details.
	}
	\label{side2}
\end{figure}

Figure~\ref{side2} shows the $R_{xy}$ distribution for the control region after applying all the selection criteria, where the MC events are normalized to the 
luminosity and a small correction for the muon misidentification is applied. A total of 505 events are observed in the control region with a background expectation of 
477$\pm$23 events, where the uncertainty includes all uncertainties discussed in section IV~B. The background is dominated by 
$\tau^- \rightarrow \pi^- \pi^+ \pi^- (\pi^0) \nu_{\tau}$ decays. The data are well reproduced by the expected background from MC within 1$\sigma$. The MC shows that the signal 
($\tau^- \to \pi^- \nu_{\tau} \mu^+ \mu^-$) contribution in this control region is small (about 2.3\%) even assuming the maximum value of the theoretical 
predictions for the branching fraction of the signal mode~\cite{Roig}. 

\section{IV. Results}

The branching fractions of $\tau^- \rightarrow \pi^-  \nu_{\tau} \ell^+ \ell^-$ decays are calculated according to the formula
\begin{eqnarray}
\mathcal{B}(\tau^- \rightarrow \pi^- \nu_{\tau} \ell^+ \ell^-) =\frac{N_{\rm obs} - N_{\rm bkg}}{ 2 \cdot \sigma_{\tau\tau} \cdot \mathcal{L}  \cdot \epsilon_{\rm sig}  },
\end{eqnarray}
where $\sigma_{\tau\tau}= (0.919~\pm~0.003)$ nb~\cite{tautau} is the cross section of $\tau\tau$ production at $\sqrt{s}=10.58$~GeV, $\mathcal{L}$ the luminosity of experimental data, $\epsilon_{\rm sig}$ the detection efficiency of signal events, $N_{\rm obs}$ the number of observed events from experimental data and $N_{\rm bkg}$ the number 
of background events.

\subsection{A. Measurement of the $\tau^{-} \rightarrow \pi^{-} \nu_{\tau} e^+ e^-$ branching fraction}

\begin{figure}[!htbp]
	\includegraphics[width=5cm,height=4cm]{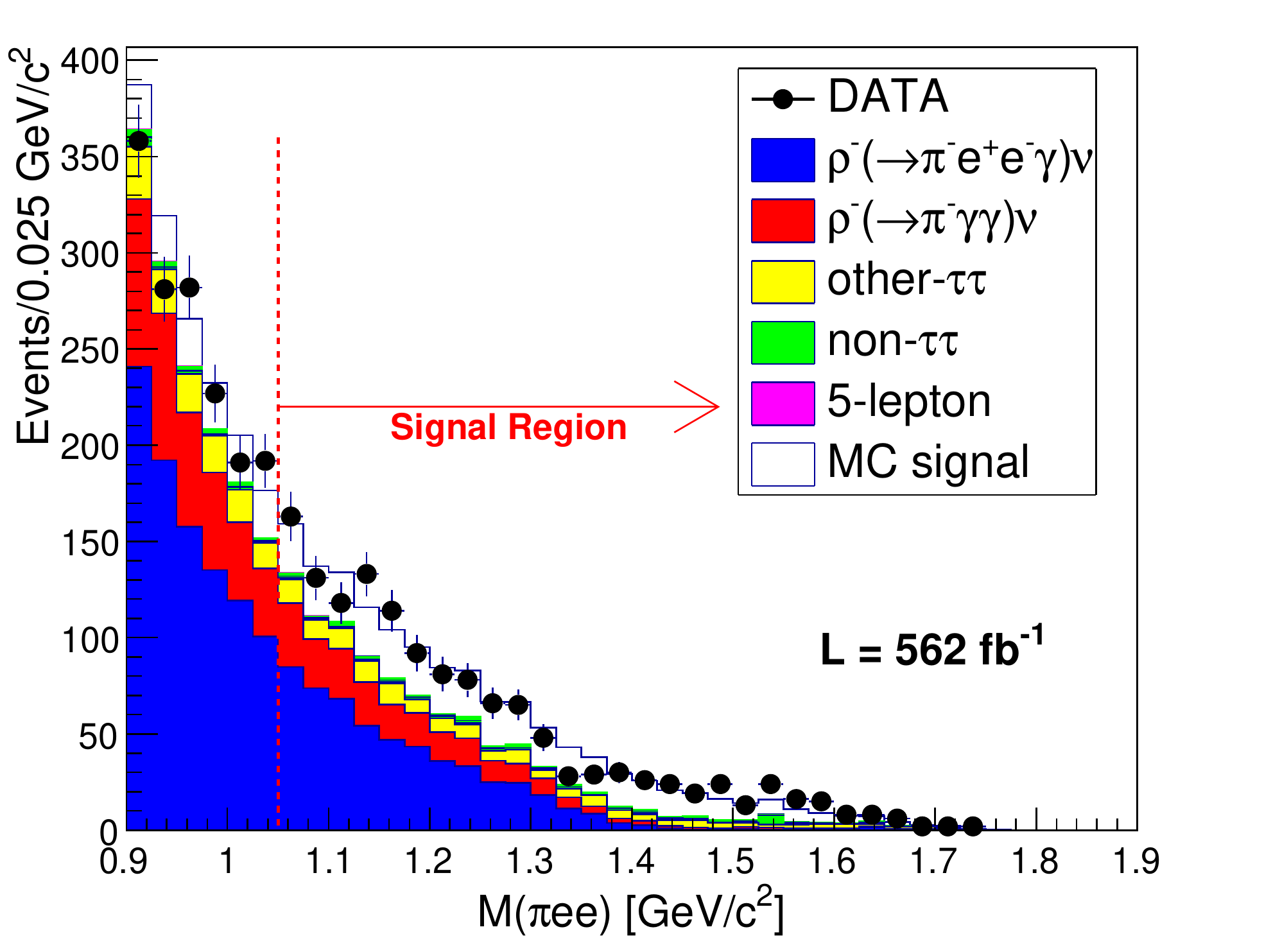} \\
	\caption{Distribution of $\pi e^+ e^-$ invariant mass ($M_{\pi ee}$) for the $\tau^- \rightarrow \pi^- \nu_{\tau} e^+ e^-$ candidates in the signal region. Circles with statistical error bars represent the experimental data. The white histogram is the expected signal using the theoretical model in Ref.~\cite{Roig}. } 
	\label{Box}
\end{figure}

The $M_{\pi e e}$ distribution of the selected sample is shown in Fig.~\ref{Box}. After applying all selection criteria, 1365 events are observed in the signal region with a background expectation of 954 $\pm$ 45 events. 
The uncertainty of background expectation takes into account all possible sources discussed later. The main background comes from the $\tau^{-} \rightarrow \pi^{-} \pi^0 \nu_{\tau}$ 
decay, in which the $\pi^0$ decays either to two photons (about 24\% of the total background), or to $e^+ e^- \gamma$ (about 56\% of the total background). Both cases are caused by one unreconstructed photon, which is common in both the signal and the control region. The MC prediction is checked by the data in the control region. 

A clear excess of data events over the background expectation is observed. The background-only hypothesis is rejected with a 7.0$\sigma$ significance. Once we select signal events in the limited $M_{\pi e e}$ mass region, 
two types of branching fractions are evaluated, the partial branching fraction in the 
limited mass region $M_{\pi e e}>1.05$~GeV/$c^2$, and the branching fraction in the full 
$M_{\pi e e}$ mass region. 
The detection efficiency in the $M_{\pi e e}>1.05$~GeV/$c^2$ region is $(6.75 \pm 0.13)$\%, where the 
dominant error is due to the model dependence of vector and axial-vector components. 
Taking this number into account, the corresponding partial branching fraction for the $M_{\pi e e}>1.05$~GeV/$c^2$ 
is obtained to be $\mathcal{B}(\tau^-\rightarrow \pi^- \nu_\tau e^+ e^-)[M_{\pi^- e^+ e^-}>1.05~{\rm GeV}/c^2] = (5.90 \pm 0.53 \pm 0.85 \pm 0.11) \times 10^{-6}$, where the first uncertainty is statistical, the second is systematic, and the third is due to the model dependence. 
The extension of this branching fraction to the full $M_{\pi e e}$ mass region depends on the assumed 
relative contribution of vector and axial-vector current of the structure-dependent terms. The detection efficiency 
of the $\tau^{-} \rightarrow \pi^{-} \nu_{\tau} e^+ e^-$ events in this case varies between two extreme cases, $(1.32 \pm 0.05)$\% 
assuming vector current only and $(2.73 \pm 0.10)$\% assuming axial-vector current only. Taking these numbers into account, we obtain 
$\mathcal{B}_{A}(\tau^-\rightarrow \pi^- \nu_\tau e^+ e^-) = (1.46 \pm 0.13 \pm 0.21) \times 10^{-5}$ 
with an assumption of pure axial-vector current and 
$\mathcal{B}_{V}(\tau^-\rightarrow \pi^- \nu_\tau e^+ e^-) = (3.01 \pm 0.27 \pm 0.43) \times 10^{-5}$ 
with pure vector current, where the first uncertainty is statistical and the second is systematic.

The systematic uncertainty of the measured branching fraction takes into account all sources and is estimated to be 14.4\% in total. The uncertainty due to the track reconstruction efficiency 
is estimated to be 4.7\% using partially reconstructed $D^{*} \rightarrow D^{0} \pi$~with~$D^0 \rightarrow \pi^- \pi^+ K^{0}_{S}$ events. The uncertainty due to particle 
identification is estimated to be 11.1\%, including contributions from the $K/\pi$ separation and lepton identification. The former is investigated with a control sample of 
$D^{*+} \rightarrow D^0 \pi^{+}_{\rm slow}, D^0 \rightarrow K^- \pi^+$ decays; the latter is studied with the $\gamma \gamma \rightarrow \ell^+ \ell^-$ and $J/\psi \rightarrow \ell^+ \ell^-$ 
processes. The uncertainty on the luminosity is obtained to be 1.4\% using Bhabha events where final-state electrons are required to be detected in the barrel region. 
Since the luminosity is used for the estimation of the number of tau-lepton pairs and the number of background events, its contribution to the uncertainty $\bigtriangleup \mathcal{B}/\mathcal{B}$ 
is 4.7\%. The uncertainty associated with the trigger efficiency is investigated by a dedicated trigger-simulation program and found to be 1.2\%. The uncertainty due to the $\pi^0$ 
reconstruction is investigated with a control sample of $\tau^- \rightarrow \pi^- \pi^0 \nu_{\tau}$, $\pi^0 \rightarrow e^+ e^- \gamma$ events and estimated to be 1.9\%. The uncertainty 
arising from the limited size of MC samples for the study of background contamination and efficiency of signal events is estimated to be 3.7\% via its binomial variation. The uncertainties 
of the branching fractions of the background modes are also taken into account and found to be 4.4\%. Finally, the uncertainty of the $\tau\tau$ production cross section at $\Upsilon(4S)$, 
0.3\%, is also included.

\subsection{B. Upper limit of the $\tau^- \rightarrow \pi^- \nu_{\tau} \mu^+ \mu^-$ branching fraction}

\begin{figure}[!htbp]
	\includegraphics[width=5cm,height=4cm]{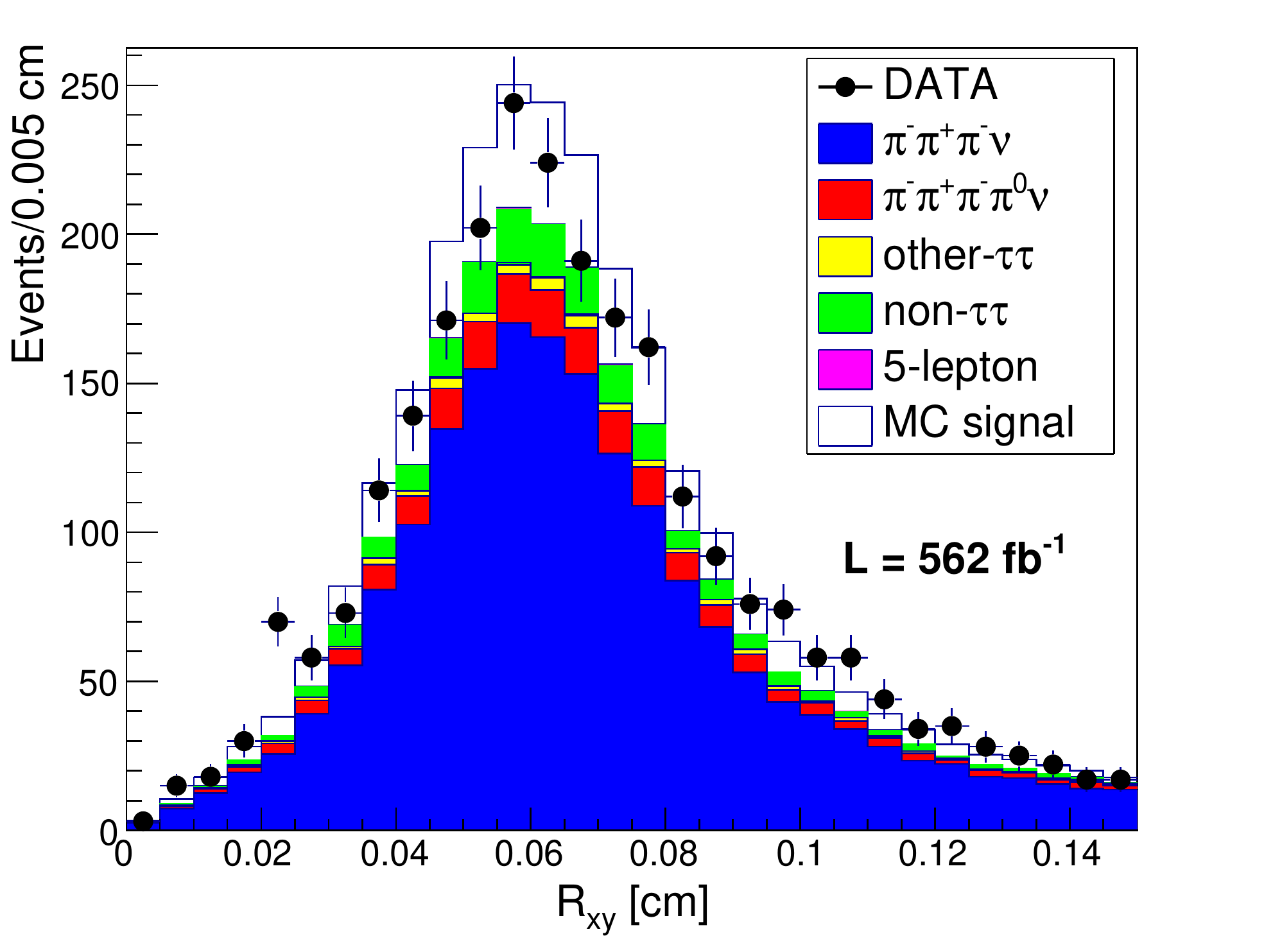} \\
	\caption{Distribution of transverse position of the $\mu^+ \mu^-$ vertex $({\rm R_{xy}})$ for the $\tau^- \rightarrow \pi^-  \nu_{\tau} \mu^+ \mu^-$ candidates. Circles with statistical error bars represent the experimental data. The white histogram is the expected signal assuming $\mathcal{B} = 1.0 \times 10^{-5}$.}
	\label{Box2}
\end{figure}

The $R_{xy}$ distribution of the selected samples is shown in Fig.~\ref{Box2}.
After applying all selection criteria, 2578 events are observed in the signal region  with a background expectation of 2244 $\pm$ 109 events. The main backgrounds are from the $\tau^{-} \rightarrow  \pi^{-} \pi^{+} \pi^{-} \nu_{\tau}$ decay (81.9\% of the total background)
and $\tau^{-} \rightarrow \pi^{-} \pi^{+} \pi^{-} \pi^0 \nu_{\tau}$ decay (8.3\% of the total background). 
The total systematic uncertainty of the background expectation is estimated to be 4.9\% including contributions from tracking (1.4\%), particle identification (3.7\%), luminosity (1.4\%), trigger (0.3\%), 
MC sizes (1.7\%), accuracies of branching fractions of background modes (1.0\%), $\tau\tau$ cross section (0.3\%), and $\pi \rightarrow \mu$ misidentification calibration (1.5\%). 
The $\pi \rightarrow \mu$ misidentification 
is determined by a control sample of $\tau^{\pm} \rightarrow \pi^{\pm} \pi^- \pi^+ \nu_{\tau}$ events. 
As a result, the excess (signal yield) is 334 $\pm$51$\pm$109 with a significance of 2.8$\sigma$, where the first (second) uncertainty is statistical (systematic). 
We determine the upper limit for the branching fraction of this decay. 

The detection efficiency of the $\tau^{-} \rightarrow \pi^{-} \nu_{\tau} \mu^+ \mu^-$ events is ($4.14 \pm 0.16$)\%. Taking into account the systematic uncertainty of the background 
expectation and the statistical uncertainty of the observed data events, an upper limit at 90\% confidence level (CL) is set on this decay mode to be $\mathcal{B}(\tau^- \rightarrow \pi^- \nu_{\tau} \mu^+ \mu^-) < 1.14 \times 10^{-5}$.

\section{V. Conclusion}
Using 562 $\rm fb^{-1}$ of data collected with the Belle detector at the KEKB asymmetric-energy $e^+e^-$ collider, the $\tau^- \rightarrow \pi^- \nu_{\tau} e^+ e^-$ 
decay is observed for the first time with 7.0$\sigma$ significance to reject the background-only hypothesis.
The partial branching fraction for the mass region $M_{\pi e e}>1.05$~GeV/$c^2$ is measured to be 
$\mathcal{B}(\tau^-\rightarrow \pi^- \nu_\tau e^+ e^-)[M_{\pi^- e^+ e^-}>1.05~{\rm GeV}/c^2] = (5.90 \pm 0.53 \pm 0.85 \pm 0.11) \times 10^{-6}$, 
where the first uncertainty is statistical, the second is systematic, and the third is due to the model dependence. 
The extension of the branching fraction to the full phase space depends on the assumed model and ranges 
from $\mathcal{B}_{A}(\tau^-\rightarrow \pi^- \nu_\tau e^+ e^-) = (1.46 \pm 0.13 \pm 0.21) \times 10 ^{-5}$ to $\mathcal{B}_{V}(\tau^-\rightarrow \pi^- \nu_\tau e^+ e^-) = (3.01 \pm 0.27 \pm 0.43) \times 10 ^{-5}$, where 
the former corresponds to a model with pure axial-vector current and the latter with pure vector current. This is the smallest decay rate of tau lepton determined to date.
An upper limit is set on the branching fraction of $\tau^- \rightarrow \pi^- \nu_{\tau} \mu^+ \mu^-$ to be $\mathcal{B}(\tau^- \rightarrow \pi^- \nu_{\tau} \mu^+ \mu^-) < 1.14 \times 10^{-5}$, 
at 90\% CL. The obtained results are consistent with the theoretical predictions~\cite{Roig} and can help to constrain relevant form factors.

\section{Acknowledgements} 
The authors would like to thank Pablo G. Roig and Gabriel L. Castro
from CINVESTAV for the extensive help and fruitful discussions.
We thank the KEKB group for the excellent operation of the
accelerator; the KEK cryogenics group for the efficient
operation of the solenoid; and the KEK computer group, and the Pacific Northwest National
Laboratory (PNNL) Environmental Molecular Sciences Laboratory (EMSL)
computing group for strong computing support; and the National
Institute of Informatics, and Science Information NETwork 5 (SINET5) for
valuable network support.  We acknowledge support from
the Ministry of Education, Culture, Sports, Science, and
Technology (MEXT) of Japan, the Japan Society for the 
Promotion of Science (JSPS), and the Tau-Lepton Physics 
Research Center of Nagoya University; 
the Australian Research Council including grants
DP180102629, 
DP170102389, 
DP170102204, 
DP150103061, 
FT130100303; 
Austrian Science Fund (FWF);
the National Natural Science Foundation of China under Contracts
No.~11435013,  
No.~11475187,  
No.~11521505,  
No.~11575017,  
No.~11675166,  
No.~11705209;  
Key Research Program of Frontier Sciences, Chinese Academy of Sciences (CAS), Grant No.~QYZDJ-SSW-SLH011; 
the  CAS Center for Excellence in Particle Physics (CCEPP); 
the Shanghai Pujiang Program under Grant No.~18PJ1401000;  
the Ministry of Education, Youth and Sports of the Czech
Republic under Contract No.~LTT17020;
the Carl Zeiss Foundation, the Deutsche Forschungsgemeinschaft, the
Excellence Cluster Universe, and the VolkswagenStiftung;
the Department of Science and Technology of India; 
the Istituto Nazionale di Fisica Nucleare of Italy; 
National Research Foundation (NRF) of Korea Grants
No.~2015H1A2A1033649, No.~2016R1D1A1B01010135, No.~2016K1A3A7A09005
603, No.~2016R1D1A1B02012900, No.~2018R1A2B3003 643,
No.~2018R1A6A1A06024970, No.~2018R1D1 A1B07047294; Radiation Science Research Institute, Foreign Large-size Research Facility Application Supporting project, the Global Science Experimental Data Hub Center of the Korea Institute of Science and Technology Information and KREONET/GLORIAD;
the Polish Ministry of Science and Higher Education and 
the National Science Center;
the Grant of the Russian Federation Government, Agreement No.~14.W03.31.0026; 
the Slovenian Research Agency;
Ikerbasque, Basque Foundation for Science, Spain;
the Swiss National Science Foundation; 
the Ministry of Education and the Ministry of Science and Technology of Taiwan;
and the United States Department of Energy and the National Science Foundation.


\begin{thebibliography}{99}
\bibitem{CC}
Unless stated otherwise, the inclusion of charge-conjugate decays is implied throughout this paper.
\bibitem{explanation}
The diagram with a $W^* \gamma^* W^*$ term is strongly suppressed by the $W$ propagator and is not shown in Fig.1.
\end{thebibliography}
\end{document}